\documentstyle[12pt,aps,prc,preprint]{revtex}
\newcommand{\be}{\begin{equation}}
\newcommand{\ee}{\end{equation}}
\newcommand{\ba}{\begin{eqnarray}}
\newcommand{\ea}{\end{eqnarray}}
\begin{document}

\begin{titlepage}
\title{\vspace*{25mm}\bf Medium effects in K$^+$ nucleus interaction from
consistent analysis of integral and differential cross sections}
\vspace{6pt}

\author{ E.~Friedman$^a$, A.~Gal$^a$
        and J.~Mare\v{s}$^b$               \vspace{6pt} \\
$^a$ Racah Institute of Physics, The Hebrew University, Jerusalem 91904,
Israel \\
$^b$ Nuclear Physics Institute, 25068 \v{R}e\v{z}, Czech Republic}

\vspace{6pt}
\maketitle

\begin{abstract}

Self consistency in the analysis of transmission measurements for K$^+$
on several nuclei in the momentum range of 500-700 MeV/c is achieved with
a `$t_{eff}(\rho)\rho$'
potential and new results are derived for total cross sections.
The imaginary part of the $t_{eff}$ amplitude is found to
increase linearly with the {\it average} nuclear density in excess of
a threshold value of 0.088$\pm$0.004~fm$^{-3}$.
This phenomenological density dependence of the K$^+$ nucleus optical
potential also gives rise to good agreement with recent measurements
of differential cross sections for elastic scattering of 715 MeV/c
K$^+$ by $^6$Li and C.
\newline 
E-MAIL: elifried@vms.huji.ac.il, avragal@vms.huji.ac.il, mares@ujf.cas.cz

\end{abstract}
\end{titlepage}

Total cross sections for the interaction of 500-700 MeV/c
K$^+$ with several nuclei were derived from transmission
experiments performed in recent years
\cite{mar90,kra92,saw93,weis94}.  Since in this energy range the KN
interaction is relatively weak and
does not vary strongly with energy,  one expects that
optical potentials close to the `$t\rho$' approximation
 will be adequate. Consequently such potentials
indeed were used for extracting
total cross sections from the data \cite{mar90,kra92,saw93,weis94}.
However, attempts to reproduce these total cross sections with optical
potentials, including in addition various degrees of conventional
nuclear physics sophistication beyond the
$t\rho$ starting point, invariably gave rise \cite{chen92} to calculated
values smaller than the measured cross sections. This as well as earlier
indications led
to speculations about nonconventional medium modifications
of the KN interaction\cite{sieg84,bro88,cai92,JKo92,GRN95}.
Very recently the same transmission experiments \cite{weis94}
were re-analyzed \cite{FGW96} to yield reaction
cross sections, for the first time, and also revised values
for the total cross sections.

The derivation of reaction
($\sigma_R$) and total ($\sigma_T$) cross sections from
transmission measurements involves the use of an
optical potential $V_{opt}$.  An obvious question of self consistency is
whether these derived $\sigma_R$ and $\sigma_T$ values
are consistent with values predicted by $V_{opt}$, or more generally, do
optical potentials which are constructed to fit $\sigma_R$ and $\sigma_T$
values lead to the same values (within errors) when used in a re-analysis
of transmission experiments.
The answer, unfortunately, has been negative so far.
As an example we quote results
for Si at 714~MeV/c; the derived reaction and total
cross sections are \cite{FGW96} 317.5$\pm$3.6 and 396.5$\pm$2.3~mb
respectively whereas the corresponding calculated values, based
on a  $t\rho$ potential,  are 253 and 328 mb respectively.
Attempting to fit the data for $^6$Li, C, Si
and Ca with $t_{eff}\rho$, or with density-dependent
potentials that include higher powers of $\rho$,
generally reduces the values of $\chi^2$ per point from
a few hundreds to a few tens.
However, if any of these modified potentials is used
 in reanalyses of the transmission
measurements then the process {\it diverges}.
It is therefore obvious that the question of
self consistency is  essential to the derivation of
integral cross sections from transmission measurements and
to the proper handling of systematic  errors.
In the present work we report on an empirical specific density
dependent potential that
achieves excellent fits to the measured reaction and total
cross sections within a  fully consistent picture. In addition
it fits well recent measurements \cite{Mic96}
of angular distributions for the elastic
scattering of 715 MeV/c K$^+$ by $^6$Li and C.

The interaction of K$^+$ with nuclei is described in the present
work by the Klein Gordon equation

\be \label{KG1}
\left[ \nabla^2 + k^2 - (2\varepsilon^{(A)}_{red}V -
V^2_c)\right] \psi = 0~~ ~~(\hbar = c = 1)
\ee

\noindent
where $k$ and $\varepsilon^{(A)}_{red}$ are the
wave number and reduced energy respectively in the $c.m.$
system, $(\varepsilon^{(A)}_{red})^{-1}=E_p^{-1}+E_t^{-1}$ in
terms of the $c.m.$ energies for the projectile and target
particles, respectively.
 $V_c$ is the Coulomb potential due to the charge
distribution of the nucleus, and $V = V_c + V_{opt}$.
The simplest possible $t\rho$ form was chosen for the optical
potential, namely

\be \label{potl}
2 \varepsilon^{(A)}_{red}V_{opt}(r)=-4\pi F_k
b_0\rho(r) ,~~ ~~ F_k = \frac{M_A\sqrt s}{M(E_t+E_p)}~~,
\ee

\noindent
where  $F_k$ is a kinematical
factor resulting from the
transformation of amplitudes between the KN and the
K$^+$ nucleus $c.m.$ systems and $b_0$ is the value of the KN scattering
amplitude in the forward direction.
$M$ is the free nucleon mass,  $M_A$ is the
mass  of the  target nucleus,
$\sqrt s$ is the total projectile-nucleon energy
in their $c.m.$ system
and the nuclear density
distribution $\rho(r)$ is normalized to $A$, the number of
nucleons in the target nucleus.
Another choice of kinematics and wave equation will be
mentioned below.

It was shown in \cite{FGW96} that by comparing the ratios
of measured to calculated $\sigma_R$ and $\sigma_T$ values
(based on $t\rho$ potentials) for C, Si and Ca to the corresponding
ratios for $^6$Li, a clear dependence on the nucleus emerges
that appears to be independent of beam momentum.
An alternative way of displaying the dependence of $\sigma_R$
and $\sigma_T$ on the nuclear medium, beyond the obvious dependence
on $A$ and $\rho$ implied by eqs.(\ref{KG1})  and (\ref{potl}), is to fit
the parameter $b_0$ of  $V_{opt}$ to the $^6$Li
data and then use this potential to {\it calculate} reaction and
total cross sections for the other, much denser nuclei. Figure \ref{ratios}
shows ratios of experimental to calculated integral cross sections
for C, Si and Ca. These ratios
deviate considerably from the value of one in a way which is
largely independent of the beam momentum.

The choice of as light a nucleus as $^6$Li to serve as a basis
for studying K meson nuclear medium effects in  terms of
optical potentials might appear questionable. Nevertheless,
the use of a potential can be bypassed at the relatively high
energies encountered here by using Glauber's eikonal
multiple scattering expansion. For example,
in very light nuclei the bulk contribution to the total
cross section comes from single scattering, i.e. $\sigma_T
\sim A\sigma$, where $\sigma$ is the KN total cross section.
For the double scattering contribution at 714 MeV/c we
calculate for $^6$Li a value of $-5\pm2$ mb compared to the
single scattering contribution of 84~mb. The uncertainty of
$\pm2$ mb reflects uncertainty in the value of $b_0$
(particularly its real part) and modifications due to nuclear
center of mass motion and Pauli and shell model correlations
which provide the longest range two-body nuclear correlations.
 The neglect of $1/A$ corrections involved in
using an eikonalized optical potential to approximate the
multiple scattering expansion does not apply to the dominant
single scattering term. Consequently it  amounts to only about 1 mb
in the present case.
Thus we estimate that the use of an optical
potential at these energies for forward K$^+$ $^6$Li
scattering does not introduce errors larger than about 2\%.
This is considerably smaller than the medium effects in the heavier
nuclei discussed in the present work.

Since integral cross sections are sensitive particularly
to the imaginary part of the optical potential, we have
tried in a first step to see whether a modification of the imaginary
part of a $t_{eff}\rho$ optical potential, via multiplication by a rescaling
factor,  separately for each target nucleus,
is capable of providing good fits to the data.
 Excellent agreement with the C, Si and Ca data was
achieved when the imaginary potential was increased by 17-25\%,
depending on the nucleus,
on top of the increase of 5-15\%
in $ImV_{opt}$, with respect to the free KN interaction,
needed to fit the $^6$Li data
 as described above.
Moreover, this target-dependent rescaling
factor turned out to be independent of the beam momentum.
Next we  proceeded to replace this {\it ad hoc} approach by
a more general model that  depends on  parameters
of the target nuclei.
We failed to reach good agreement with the data by using
the local density in modifying the potential. However,
it was possible to achieve good fits to the data by correlating
 $ImV_{opt}$ with the {\it average} nuclear
density, defined as follows:

\be \label{rhobar}
\overline{\rho}=\frac{1}{A}\int\rho^2d{\bf r}.
\ee

\noindent
Values of $\overline{\rho}$ can be obtained quite reliably from
Hartree-Fock calculations or from simpler single-particle
calculations \cite{BGr69,MiH73} that are constrained by  r.m.s.
radii of nuclear charge distributions.
The results for the nuclei considered here
are 0.049, 0.103, 0.110 and 0.107~fm$^{-3}$ for $^6$Li, C, Si and Ca
respectively, with an uncertainty of about $\pm$1\% for the last
three cases. The enhancement of $ImV_{opt}$ was then assumed to
take place above a threshold average density $\rho_c$, by multiplying
the imaginary part of the
conventional potential eq.(\ref{potl}) by the rescaling factor
$1+\beta(\overline{\rho}-\rho_c)\Theta(\overline{\rho}-\rho_c)$,
and a fit was made at each momentum
varying the  parameters $b_0$, $\beta$ and $\rho_c$.
The last two parameters were found to have the same values at all
four beam momenta.
Table
\ref{trhobar} shows that excellent fits  are
obtained  within this linear rescaling model.
In contrast, attempting to replace the
threshold function $\Theta(\overline{\rho}-\rho_c)$ by a power of
the average density failed to produce fits to the data.
 The average values of $\rho_c$=0.088$\pm$0.004~fm$^{-3}$
and $\beta$=13.0$\pm$3.4 fm$^3$ were used in these calculations.
This value of $\rho_c$ is considerably larger than
$\overline{\rho}$($^6$Li) so that the $^6$Li data
are fitted, as above, merely by adjusting the parameter $b_0$.
Note that the inequality $\overline{\rho}$(Ca)$>\overline{\rho}$(C)
(see above), which implies stronger rescaling of $ImV_{opt}$ for
Ca than for C, does not translate into larger
$\sigma$(exp.)/$\sigma$(calc.) values (see fig.\ref{ratios}) for
Ca than for C since multiple scattering  is more effective
in the former. The departure of the fitted (medium) KN forward scattering
amplitude $b_0$ from its free-space value (given in parentheses
in table \ref{trhobar}) is worth a comment. $Imb_0$ is found to gradually
increase with energy up to 16\% above its free-space value. This
increase is quantitatively consistent with the estimates of
Garcia-Recio {\it et al.} \cite{GRN95} of the contribution due to
meson exchange currents. However, we have no clues for the theoretical
significance of $\rho_c$ in providing a threshold density above
which new reaction channels in the K-nucleus interaction open up,
beyond those accounted for by the reactive content of the
impulse approximation. As for $Reb_0$, we observe from table \ref{trhobar}
a clear tendency for {\it less} repulsive KN interaction with
increasing energy than that implied by the free space values.
Such additional {\it attraction} could be expected due to the
proximity of the K$^*$N and K$\Delta$ channels.

The next question is that of self consistency. With the
above enhancement of $ImV_{opt}$
 we re-analyzed all the transmission measurements
and obtained new values for $\sigma_R$ that almost agree, within
errors, with the previous values based on a $t\rho$ potential \cite
{FGW96}. Values of $\sigma_T$ are higher a little more
 (3-5\%) than the
 previous ones \cite{FGW96},
as expected  on grounds of the extent of theoretical
input involved in the analysis of transmission
measurements \cite{FGW96,Ari93}.
Fits to these new $\sigma_R$
and $\sigma_T$ values  left the parameters of
table \ref{trhobar} essentially
unchanged and another round of re-analysis of
the transmission measurements showed that full convergence had been
achieved, at  all four beam momenta. The parameters in table \ref{trhobar}
correspond to the converged, self consistent values of $\sigma_R$
and $\sigma_T$.

A further test of the whole procedure is now possible at the
highest momentum thanks to the very recent
publication of experimental differential cross sections for K$^+$
elastic scattering at
715 MeV/c  on $^6$Li and C \cite{Mic96}. These angular
distributions were therefore used in optical model fits, with a variety
of potentials including the one of table \ref{trhobar}.
The normalization of the data was allowed to vary too, within the quoted
range of $\pm$15\%. The best fits to the elastic scattering data without
rescaling  $ImV_{opt}$ achieved reasonable fits to the angular
distributions but with $\chi^2$ per point of about 20 for the
reaction and total cross sections (for $^6$Li, C, Si and Ca) at this
momentum.
In contrast, when  $ImV_{opt}$ was rescaled as in table \ref{trhobar},
 values of $\chi^2$ per point for the reaction and total cross
sections were in the range of 1.6 to 2.0. The normalization
factor for the angular distribution data
was well determined at the acceptable value of
0.92$\pm$0.05. Figure \ref{angdist} shows the good agreement between
calculated and measured angular distributions. The potential of
fig.\ref{angdist} was then used in a further re-analysis of
the transmission
measurements at 714 MeV/c to yield yet another set of reaction and total
cross sections, based this time on potentials that fit also
the available angular distributions. Table~\ref{final} summarizes
$\sigma_R$ and $\sigma_T$ values as obtained from transmission measurements
using (a) the $t\rho$ potential (eqs.(\ref{KG1}) and (\ref{potl}))
and (b), (c) two potentials
incorporating
the $\Theta(\overline{\rho}-~\rho_c)$
rescaling, where (c) is also
constrained by fits to differential cross sections for  elastic
scattering. For (b) and (c) the process converges, which
is not the case for (a). Values of $\sigma_T$ are larger by 3-5\%
than those found in non self-consistent analyses. Values of $\sigma_R$
remain practically unchanged.

One final comment on the kinematics and wave equation is in order.
Since there is no agreed relativistic wave equation to be used with
a finite mass target once a potential is introduced, we have checked
the dependence of our results on the wave equation and kinematics
by repeating all the calculations with the Goldberger-Watson
equation \cite{Gol64}. Details do change but the overall picture
remains unchanged. In particular, there is a threshold average
nuclear density of about 0.09 fm$^{-3}$
above which the imaginary potential must be rescaled
in order to achieve agreement between calculated and measured
integral cross section within a consistent analysis.
A major advantage of using the wave equation (\ref{KG1}) with
the potential (\ref{potl}) is that it yields in the eikonal
approximation an attenuation factor which agrees with the
semiclassical expression ($\sigma\rho)^{-1}$
for the mean free path.  This ensures the correct reactive
content of the procedure adopted here.

To summarize, empirical rescaling of the imaginary part of the
optical potential with the {\it average nuclear density in excess of
a threshold
value} solves the long standing problem of inconsistencies in the
extraction of reaction and total cross sections from
transmission measurements for the interaction
of 500-700 MeV/c K$^+$ with nuclei. The self consistent values of
total cross sections are 3-5\% larger than previously published values.
This potential is also very successful
in reproducing the available angular distributions of elastically
scattered K$^+$ by $^6$Li and C. The empirical finding that the absorptive
part of $V_{opt}$ increases  with the average
nuclear density above a threshold value $\rho_c$,
may signal that novel degrees of freedom are excited in K$^+$ nucleus
interactions.

This research was partially supported by the Israel Science
Foundation administered by the Israel Academy of Sciences
and Humanities (E.F.) and by the U.S.-Israel Binational Science
Foundation (A.G.). A.G. would like to thank B.K. Jennings for his
hospitality at TRIUMF, and J.M. acknowledges the hospitality of the
Racah Institute of Physics at the Hebrew University.

\begin{figure}
\caption{Ratios between experimental and calculated cross
sections for calculations based on fits to $^6$Li.
Full squares represent $\sigma_R$ and open circles represent
$\sigma_T$.}
\label{ratios}
\end{figure}
\begin{figure}
\caption{Fits to angular distributions for elastic scattering
of 715 MeV/c K$^+$ by $^6$Li and C (see text).} \label{angdist}
\end{figure}

\begin{table}
\caption{Fits to K$^+$ nucleus $\sigma_R$ and $\sigma_T$ values
obtained by rescaling $ImV_{opt}$ by the factor
 $1+\beta(\overline{\rho}-\rho_c)\Theta(\overline{\rho}-\rho_c)$
with $\rho_c$=0.088 fm$^{-3}$, $\beta$=13.0 fm$^3$.
Values in parentheses are for the free KN interaction.}
 \label{trhobar}
\begin{tabular}{|cccc|}
p(MeV/c)&$Reb_0$(fm)&$Imb_0$(fm)&
$\chi^2$/N \\ \hline
488&-0.154$\pm$0.012&0.160$\pm$0.002&0.6\\
 &(-0.178)&(0.153)&  \\
531&-0.119$\pm$0.012&0.186$\pm$0.002&1.2\\
 &(-0.172)&(0.170) & \\
656&-0.035$\pm$0.062&0.241$\pm$0.002&0.2\\
 &(-0.165)&(0.213)&   \\
714&-0.044$\pm$0.064&0.265$\pm$0.001&1.0\\
 &(-0.161)&(0.228)&   \\
\end{tabular}
\end{table}
\newpage
\begin{table}
\caption{K$^+$ nucleus reaction and total cross sections
derived from transmission measurements at 714MeV/c,
based on potentials (a), (b) and (c) (see text).} \label{final}
\begin{tabular}{|ccccccccc|}
potl. &\multicolumn{4}{c}{$\sigma_R$ (mb)}&&\multicolumn{2}
{c}{$\sigma_T$ (mb)}& \\
 &Li&C&Si&Ca&Li&C&Si&Ca \\ \hline
(a)&80.7$\pm$1.2&151.8$\pm$1.5&318.7$\pm$3.6&413.7$\pm$5.5&
  86.8$\pm$0.6&177.4$\pm$0.9&392.0$\pm$2.3&523.2$\pm$2.8 \\
(b)&82.2$\pm$1.2&152.8$\pm$1.5&320.2$\pm$3.6&417.1$\pm$5.5&
88.5$\pm$0.6&183.8$\pm$0.9&411.3$\pm$2.3&550.4$\pm$2.8 \\
(c)&82.1$\pm$1.2&150.8$\pm$1.5&317.3$\pm$3.6&416.8$\pm$5.5&
89.1$\pm$0.6&183.1$\pm$0.9&411.9$\pm$2.3&554.7$\pm$2.8 \\
\end{tabular}
\end{table}

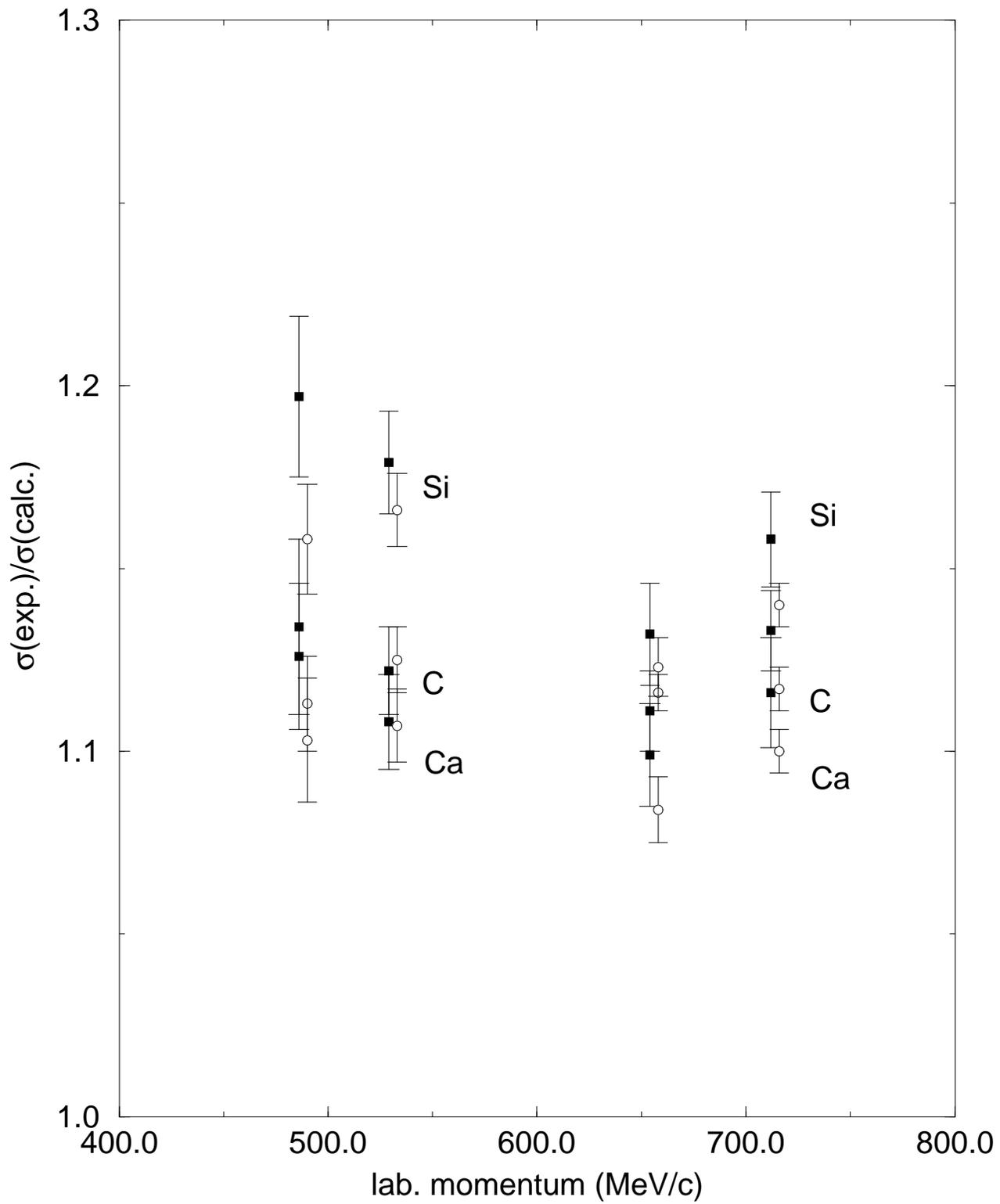
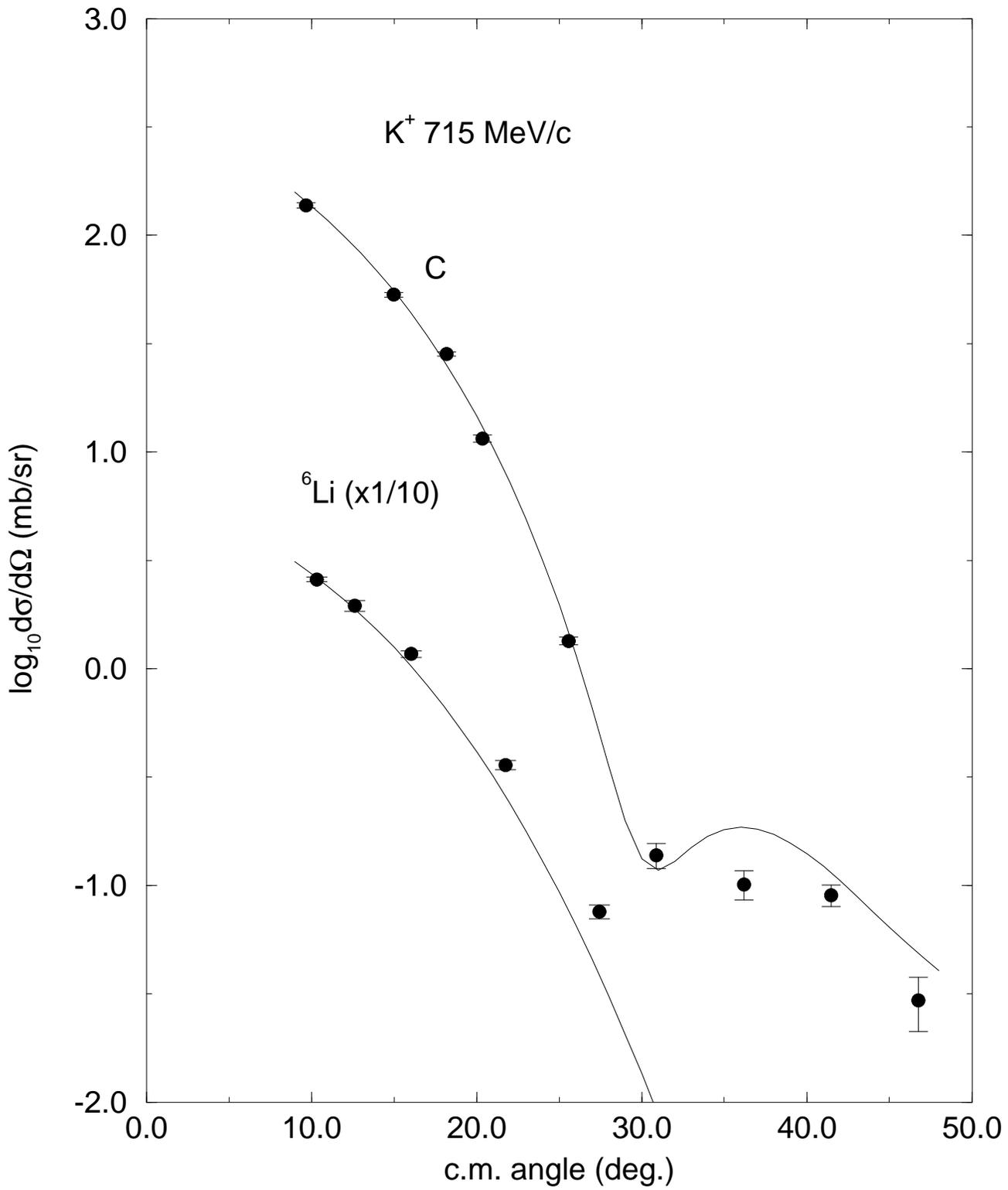
\end{document}